\documentclass[twocolumn,showpacs,amsmath,amssymb,superscriptaddress]{revtex4-1}
\usepackage{dcolumn}
\usepackage{color}
\usepackage{graphicx}
\usepackage{amsmath}
\usepackage{multirow}
\usepackage{bm}
\usepackage{url}

\newcommand\+{\dagger}

\begin{document}

\title{Description  of neutron-rich odd-mass  krypton isotopes within the 
interacting boson-fermion model based on the Gogny energy density functional
}

\author{K.~Nomura}
\affiliation{Physics Department, Faculty of Science, University of
Zagreb, HR-10000 Zagreb, Croatia}

\author{R.~Rodr\'iguez-Guzm\'an}
\affiliation{Physics Department, Kuwait University, 13060 Kuwait, Kuwait}

\author{L.~M.~Robledo}
\affiliation{Center for Computational Simulation,
Universidad Polit\'ecnica de Madrid,
Campus de Montegancedo, Boadilla del Monte, 28660-Madrid. Spain}

\affiliation{Departamento de F\'\i sica Te\'orica, Universidad
Aut\'onoma de Madrid, E-28049 Madrid, Spain}

\date{\today}

\begin{abstract}
The low-lying structure of  neutron-rich odd-mass Kr 
 isotopes is studied within the interacting boson-fermion 
 model (IBFM) based on the Gogny-D1M energy density functional (EDF). 
 The $(\beta,\gamma)$-deformation energy surfaces, spherical single-particle energies and
occupation probabilities of the odd-mass systems are 
obtained using the constrained Hartree-Fock-Bogoliubov (HFB) approximation. 
 Those quantities are used as a microscopic input to determine most of the parameters 
 of the IBFM Hamiltonian. 
 The remaining parameters are specifically tailored to the  experimental
 spectrum for each of the studied odd-mass nuclei. A gradual
 structural evolution is predicted for the odd-mass isotopes
  $^{87-95}$Kr as a function of the nucleon number which, agrees well
 with the gradual growth of collectivity observed experimentally in the
 neighboring even-even  isotopes.
\end{abstract}


\maketitle


\section{Introduction}

The neutron-rich nuclei in the mass
region $A\approx 100$ exhibit a rich variety of nuclear structure 
phenomena, such as  quantum phase transitions  \cite{cejnar2010} associated to
ground state shape evolution or the presence of  
shape  coexistence  \cite{heyde2011}. In particular, the neutron-rich Kr ($Z=36$) isotopes 
have recently attracted much attention because of their 
shape evolution with neutron number. For instance, at variance with  the 
neighboring even-even Sr ($Z=38$) and Zr ($Z=40$)
nuclei, where a rapid structural change has been suggested  around
$N=60$ \cite{togashi2016,kremer2016,clement2017}, the shape 
transition is rather moderate in the Kr
isotopic chain \cite{albers2012}. A number of
experiments, using radioactive beams, have already been 
devoted to the study and characterization of the 
properties 
of neutron-rich even-even Kr isotopes
\cite{naimi2010,albers2012,albers2013,dudouet2017,flavigny2017}. Nuclei
of this region of the nuclear chart have also been studied within 
several theoretical frameworks (see, for example, 
\cite{rayner2010odd-1,albers2012,albers2013,trodriguez2014,xiang2016,nomura2016zr,togashi2016,dudouet2017,flavigny2017,nomura2017kr}, and references therein).

Spectroscopic data are also available for neutron-rich odd-mass Kr 
nuclei like $^{91}$Kr \cite{rzacaurban2017}, $^{93}$Kr \cite{hwang2010} 
and $^{95}$Kr \cite{genevey2006}. However, from a theoretical point of 
view, those odd-mass systems have not yet been as much studied as the 
even-even ones. This is  because the description of odd-mass systems 
tends to be cumbersome, as it requires to treat  both the 
single-particle and collective degrees of freedom on an equal footing 
\cite{bohr1953}. For example, within the energy density functional 
(EDF) framework, the need to consider blocked one-quasiparticle wave 
functions \cite{RS} is what complicates the description of odd-mass 
nuclei. Those blocked wave functions break time-reversal invariance 
which requires the evaluation of time-odd fields to solve the 
corresponding Hartree-Fock-Bogoliubov (HFB) equations. Furthermore, due 
to the (nonlinear) self-consistent character of the HFB equations, 
there is no guarantee to obtain the lowest energy solution by blocking 
the quasiparticle with the lowest one-quasiparticle excitation energy 
of the neighboring even system. Therefore, the blocking procedure has 
to be repeated starting from several different low-lying one-quasiparticle 
states of the neighboring nuclei. From what has already been mentioned, 
it is clear that alternative routes should be explored to afford the 
computational cost of the EDF description of odd-mass nuclei. Within 
this context, the so called Equal Filling Approximation (EFA) 
\cite{perezmartin2008,rayner2010odd-1} has emerged as a useful tool for 
microscopic studies of odd-mass nuclei (see, for example, 
\cite{rayner2010odd-2,rayner2010odd-3}, and references therein). Though 
the HFB-EFA alleviates the computational effort required to study 
medium and heavy odd-mass nuclei, it still provides limited access to 
their spectroscopic properties.

In this paper, we consider the spectroscopy of neutron-rich odd-mass Kr 
isotopes. To this end, we have resorted to a method 
\cite{nomura2016odd} that combines the advantages of both the energy 
density functional (EDF) framework and the ones of the interacting 
boson-fermion model (IBFM) \cite{IBFM}. The deformation energy surfaces 
for the neighboring even-even nuclei as well as the single-particle 
energies and occupation probabilities for the odd-mass systems, are 
computed using  the (constrained) self-consistent mean-field (SCMF) 
approximation. Those quantities are then used as a microscopic input to 
determine most of the corresponding IBFM Hamiltonian. A few remaining 
parameters, i.e., coupling constants for the boson-fermion interaction, 
are then specifically fitted to reproduce experimental spectra in each 
odd-mass nucleus. The method allows a systematic and efficient 
description of the spectroscopic properties of odd-mass nuclei in 
various mass regions \cite{nomura2017odd-2,nomura2017odd-3}. In this 
study, we have considered  $^{87,89,91,93,95}$Kr which are the heaviest 
odd-mass Kr isotopes for which experimental data for excitation spectra 
are available. Calculations for even-even Kr isotopes, have already 
been carried out in Refs.~\cite{albers2012,albers2013,nomura2017kr} 
using an interacting boson model (IBM) based on the Gogny-EDF. It has 
been shown, that the even-even nuclei $^{86-94}$Kr undergo a gradual 
transition from nearly spherical to $\gamma$-soft and then to oblate 
shapes with increasing neutron number. Those results point to a 
moderate growth of collectivity as observed in the experimental data 
\cite{albers2012}.

Recent experimental studies as well as EDF-based calculations 
\cite{trodriguez2014,nomura2017kr} have suggested that  shape 
coexistence is likely to occur in the even-even Kr isotopes  beyond the 
mass number $A=96$. In order to study those nuclei, intruder 
configurations associated with different nuclear shapes have to be 
introduced within the  IBM formalism  \cite{albers2013,nomura2017kr}. 
However, as shown below, shape coexistence plays a minor role for  the 
even-even core nuclei $^{86-94}$Kr considered in this work and 
therefore there is no need in our  IBFM calculations to consider 
intruder configurations as it would be the case for those isotopes with  
$A\geq 96$.
 
The low-lying structure of the $N=53$ isotones $^{85}$Ge, $^{87}$Se and 
$^{89}$Kr has been studied in Ref.~\cite{alkhudair2015} within the 
IBFM-2 phenomenology. Aside from that study, and to the best of our 
knowledge, the IBFM framework has rarely been applied to the mass 
region $A\approx 100$ beyond the neutron number $N=50$. Perhaps, this 
is because  only recently experimental data  have become available for 
both even-even and odd-mass nuclei in this region. Therefore, one of 
the  aims of this study will be to test the validity of employed 
theoretical framework in the description of  neutron-rich odd-mass 
isotopes in the mass region $A\approx 100$ .

The paper is organized as follows. In Sec.~\ref{sec:model}, we briefly 
describe the  theoretical scheme used in this work. In 
Sec.~\ref{sec:even}, we review some key results for even-even Kr 
isotopes. Then in Sec.~\ref{sec:odd}, we discuss the spectroscopic 
properties for odd-mass Kr nuclei. In particular, we consider the 
systematic of the low-energy yrast levels, detailed level schemes for 
individual nuclei, $B(E2)$ and $B(M1)$ transition strengths, 
spectroscopic quadrupole and magnetic moments. Finally, 
Sec.~\ref{sec:summary} is devoted to concluding remarks and work 
perspectives.


\section{Theoretical procedure\label{sec:model}}


In this section we briefly outline the theoretical framework used 
in this study. For a more detailed account, the reader is referred to
Refs.~\cite{nomura2016odd,nomura2017odd-2}.

The IBFM
Hamiltonian $\hat H_{\rm IBFM}$ consists of the neutron-proton IBM (IBM-2) Hamiltonian
$\hat H_{\rm B}$, the single-particle fermion Hamiltonian $\hat H_{\rm F}$, and
the boson-fermion interaction term $\hat H_{\rm BF}$

\begin{eqnarray}
\label{eq:ham}
 \hat H_{\rm IBFM} = \hat H_{\rm B} + \hat H_{\rm F} + \hat H_{\rm
  BF}. 
\end{eqnarray}
The IBM-2 term  $\hat H_{\rm B}$ is comprised of the neutron (proton) $s_\nu$ and $d_\nu$ ($s_\pi$
and $d_\pi$) bosons, which are associated with 
collective pairs of valence neutrons (protons) with spin $J=0^+$ and
$2^+$, respectively \cite{OAI}, outside the $^{78}$Ni double-magic core. 
We have used the same Hamiltonian  $\hat H_{\rm B}$ as in
Refs.~\cite{albers2012,albers2013}, where  the 
$(\beta,\gamma)$-deformation energy surfaces have been computed 
using the constrained 
Hartree-Fock-Bogoliubov (HFB) method \cite{rayner2010pt} based on the 
parametrization D1M of the Gogny-EDF
\cite{Gogny,D1M}. The corresponding mean-field 
$(\beta,\gamma)$-deformation energy surfaces
are then mapped onto the expectation value of the
IBM-2 Hamiltonian \cite{nomura2008,nomura2010}.
This procedure determines the 
values of the Hamiltonian parameters (see, \cite{albers2013}).
For the fermion valence space, we have taken the $3s_{1/2}$, $2d_{3/2}$,
$2d_{5/2}$, and $1g_{7/2}$ orbitals of the whole neutron major shell
$N=50-82$ to describe positive-parity states in the studied odd-mass Kr
nuclei. Since there is no experimental information about negative-parity
states for the considered odd-mass Kr nuclei, we restrict our discussion 
to the positive-parity ones in this paper.

The $\hat H_{\rm BF}$ interaction term is taken from  \cite{IBFM}: 
\begin{eqnarray}
\label{eq:bf}
 \hat H_{\rm BF} = \Gamma_\nu\hat Q_{\pi}^{(2)}\cdot\hat q_{\nu}^{(2)} +
  \Lambda_\nu\hat V_{\pi\nu} + A_\nu\hat n_{d\nu}\hat
  n_{\nu}.
\end{eqnarray}
The first term represents the quadrupole dynamical term, 
given by the product of a strength constant $\Gamma_\nu$ times the boson
quadrupole operator for proton bosons $\hat Q^{(2)}_{\pi}$ times 
the fermion quadrupole operator for the odd neutron $q^{(2)}_\nu$. The latter
is given by  
\begin{eqnarray}
\hat
q^{(2)}_\nu=\sum_{jj'}\gamma_{jj'}(a^\+_{j\nu}\times\tilde
a_{j'\nu})^{(2)},
\end{eqnarray} 
where $\gamma_{jj'}=(u_ju_{j'}-v_jv_{j'})Q_{jj'}$. The $\gamma_{jj'}$
matrix element is the product of occupancy factors ($v_{j}$ and $u_{j}$ )
of the fermion orbitals times the matrix element of the
quadrupole operator in the single particle basis  $Q_{jj'}=\langle
j||Y^{(2)}||j'\rangle$. The second term in Eq.~(\ref{eq:bf})
is the exchange interaction with strength $\Lambda_\nu$. It accounts 
for the fact bosons are built from nucleon pairs. For the
operator $\hat V_{\pi\nu}$ we have used the expression

\begin{eqnarray}
\label{eq:exc}
 \hat V_{\pi\nu} =&& -(s_{\pi}^\+\tilde d_{\pi})^{(2)}
\cdot
\Bigg\{
\sum_{jj'j''}
\sqrt{\frac{10}{N_\nu(2j+1)}}\beta_{jj'}\beta_{j''j} \nonumber \\
&&:((d_{\nu}^\+\times\tilde a_{j''\nu})^{(j)}\times
(a_{j\nu}^\+\times\tilde s_\nu)^{(j')})^{(2)}:
\Bigg\} + (H.c.), \nonumber \\
\end{eqnarray}
where $\beta_{jj'}=(u_jv_{j'}+v_ju_{j'})Q_{jj'}$.
As can be seen from Eq.~(\ref{eq:exc}), $\hat V_{\pi\nu}$ 
should be considered 
when the neutron boson number satisfies $N_\nu\neq 0$. Therefore, it is
omitted for the semi-magic nucleus $^{87}$Kr for which $N_\nu=0$. 
The third term in Eq.~(\ref{eq:bf}) is the monopole interaction with
strength $A_\nu$. 
In the same equation, $\hat n_{d\nu}$ is the number operator for neutron $d$ bosons.
On the other hand, $n_\nu$ reads 

\begin{eqnarray}
\hat
n_{\nu}=\sum_j(-\sqrt{2j+1})(a^\+_{j\nu}\times\tilde a_{j\nu})^{(0)}. 
\end{eqnarray}

The boson-fermion Hamiltonian $\hat H_{\rm BF}$ of Eq.~(\ref{eq:bf}) can
be justified from microscopic considerations \cite{scholten1985,IBFM}:
both the quadrupole dynamical and exchange terms act predominantly
between protons and neutrons (i.e., between odd neutron and proton bosons), while the
monopole interaction acts between like-particles (i.e., between odd neutron
and neutron bosons). The single-particle energies $\epsilon_j$ and the
occupation probabilities $v^2_j$ for the odd neutron at orbital $j$, have been obtained 
via Gogny-HFB calculations constrained to zero quadrupole moment \cite{nomura2017odd-2}. 
The coupling constants
$\Gamma_{\nu}$, $\Lambda_\nu$ and $A_\nu$ have been fitted to  reproduce the lowest-lying
positive-parity levels for each of the considered odd-mass  nuclei \cite{nomura2016odd}. 
Their values, together with $\epsilon_j$ and
$v^2_2$, are  listed in Table~\ref{tab:para}. 

Let us make some remarks on the results shown in Table~\ref{tab:para}:
\begin{itemize}
	
\item For the sake of simplicity, a constant value  $\Gamma_\nu=0.80$ MeV 
      has been used for $^{89-95}$Kr. We have verified, that the calculated 
      excitation energies are not very sensitive to this parameter. 

\item The $\Lambda_\nu$ values for $^{93,95}$Kr are roughly a factor
      three larger than those for $^{89,91}$Kr. This reflects that the 
      ground state's spin changes from $^{91}$Kr to $^{93}$Kr 
      \cite{rzacaurban2017,data} (see also, Fig.~\ref{fig:odd}).

\item For  $^{91,93,95}$Kr, the monopole term was not introduced for the
      $2d_{5/2}$ orbital. This is because the SCMF single-particle energy 
      of the $2d_{5/2}$ orbital turned out to be too low ($\approx$ 1.7 MeV) 
      as compared to the one for the $3s_{1/2}$ orbital. 

\end{itemize}

\begin{table}
 \begin{center}
\caption{\label{tab:para} The single-particle energies $\epsilon_j$ (in MeV units)
  and   occupation probabilities $v^2_j$ of odd neutron at each orbital,
  and the fitted strength parameters $\Gamma_\nu$, 
  $\Lambda_\nu$ and $A_\nu$ (in MeV units) employed in the present study
  for the odd-mass nuclei $^{87-95}$Kr. Note that the monopole term has
  not been introduced for the $2d_{5/2}$ orbital. For details, see the
  main text.}
  \begin{tabular}{lcccccccc}
\hline\hline
 & & $3s_{1/2}$ & $2d_{3/2}$ & $2d_{5/2}$ & $1g_{7/2}$ & $\Gamma_\nu$ &
   $\Lambda_\nu$ & $A_{\nu}$ \\ 
\hline
\multirow{2}{*}{$^{87}$Kr} & $\epsilon_j$ & 1.766 & 2.589 & 0.000 & 2.753 & \multirow{2}{*}{0.56} & \multirow{2}{*}{--} & \multirow{2}{*}{--} \\
 & $v^2_j$ & 0.011 & 0.009 & 0.152 & 0.007 & & & \\[0.5em]
\multirow{2}{*}{$^{89}$Kr} & $\epsilon_j$ & 1.702 & 2.722 & 0.000 & 2.744 & \multirow{2}{*}{0.80} & \multirow{2}{*}{0.87} & \multirow{2}{*}{0.0} \\
 & $v^2_j$ & 0.042 & 0.025 & 0.443 & 0.022 & & & \\[0.5em]
\multirow{2}{*}{$^{91}$Kr} & $\epsilon_j$ & 1.662 & 2.830 & 0.000 & 2.720 & \multirow{2}{*}{0.80} & \multirow{2}{*}{0.5} & \multirow{2}{*}{-1.2} \\
 & $v^2_j$ & 0.109 & 0.041 & 0.705 & 0.045 & & & \\[0.5em]
\multirow{2}{*}{$^{93}$Kr} & $\epsilon_j$ & 1.672 & 2.885 & 0.000 & 2.657 & \multirow{2}{*}{0.80} & \multirow{2}{*}{2.4} & \multirow{2}{*}{-2.5} \\
 & $v^2_j$ & 0.283 & 0.080 & 0.846 & 0.116 & & & \\[0.5em]
\multirow{2}{*}{$^{95}$Kr} & $\epsilon_j$ & 1.697 & 2.883 & 0.000 & 2.554 & \multirow{2}{*}{0.80} & \multirow{2}{*}{2.4} & \multirow{2}{*}{-1.8} \\
 & $v^2_j$ & 0.463 & 0.150 & 0.882 & 0.240 & & & \\[0.5em]
\hline\hline
  \end{tabular}
 \end{center}
\end{table}

The diagonalization of the IBFM-2 Hamiltonian provides both  excitation energies
and wave functions. Those IBFM-2 wave functions have been employed to computed the $B(E2)$ and
$B(M1)$ transition rates as well as the  spectroscopic quadrupole $Q(J)$ and
magnetic $\mu(J)$
moments. The E2/M1 transition operator is the sum of a boson and a fermion part
 \begin{eqnarray}
 \hat T^{(E2/M1)} = \hat T_{\rm B}^{(E2/M1)} + \hat T_{\rm F}^{(E2/M1)}. 
\end{eqnarray}
For the boson E2 operator we have assumed  the form 
$\hat T^{(E2)}_{\rm B} = e_{\rm B}^\nu \hat Q_\nu + e_{\rm B}^\pi \hat Q_\pi$. 
The   fermion operator $\hat T^{(E2)}_{\rm F}$ is defined in Ref.~\cite{nomura2017odd-2}. 
For the proton boson effective charge $e_B^\pi$ we have used   the value
$e_{\rm B}^\pi=0.07\,e$b so as to reproduce the 
experimental $B(E2; 2^+_1\rightarrow 0^+_1)$ transition probability for the semi-magic
nucleus $^{86}$Kr. We have taken  the neutron boson charge  $e_{\rm B}^\nu$ to be $e_{\rm B}^\nu=0.035\,e{\rm
b}=e_{\rm B}^\pi/2$ in analogy with the  shell-model effective charges that satisfy 
${e^\pi}/{e^\nu}\approx 2$. On the other hand, for the neutron effective charge, we have 
considered the  value $e_{\rm F}^\nu=0.5\,e$b.  The M1 operator for the boson system reads 
\begin{eqnarray}
 \hat T^{(M1)}_{\rm B} = \sqrt{\frac{3}{4\pi}}(g_{\rm B}^\nu \hat L_\nu +
g_{\rm B}^\pi \hat L_\pi),
\end{eqnarray}
with $g_{\rm B}^\tau$ and $\hat L_{\tau}$ ($\tau=\nu,\pi$) being the
neutron/proton boson $g$-factor and angular momentum operator,
respectively. 
We have employed the standard values, $g_{\rm B}^\nu=0\,\mu_N$ and $g_{\rm B}^\pi=1.0$ $\mu_N$. 
The fermion M1 operator $\hat T^{(M1)}$ is defined in
Ref.~\cite{nomura2017odd-2}. 
For the fermion $g$-factors, we have adopted  $g_l=0\,\mu_N$, and the free value of 
$g_s$ has been quenched by 30 \%.



\section{Results for even-even Kr nuclei\label{sec:even}}



\begin{figure}[htb!]
\begin{center}
\includegraphics[width=\linewidth]{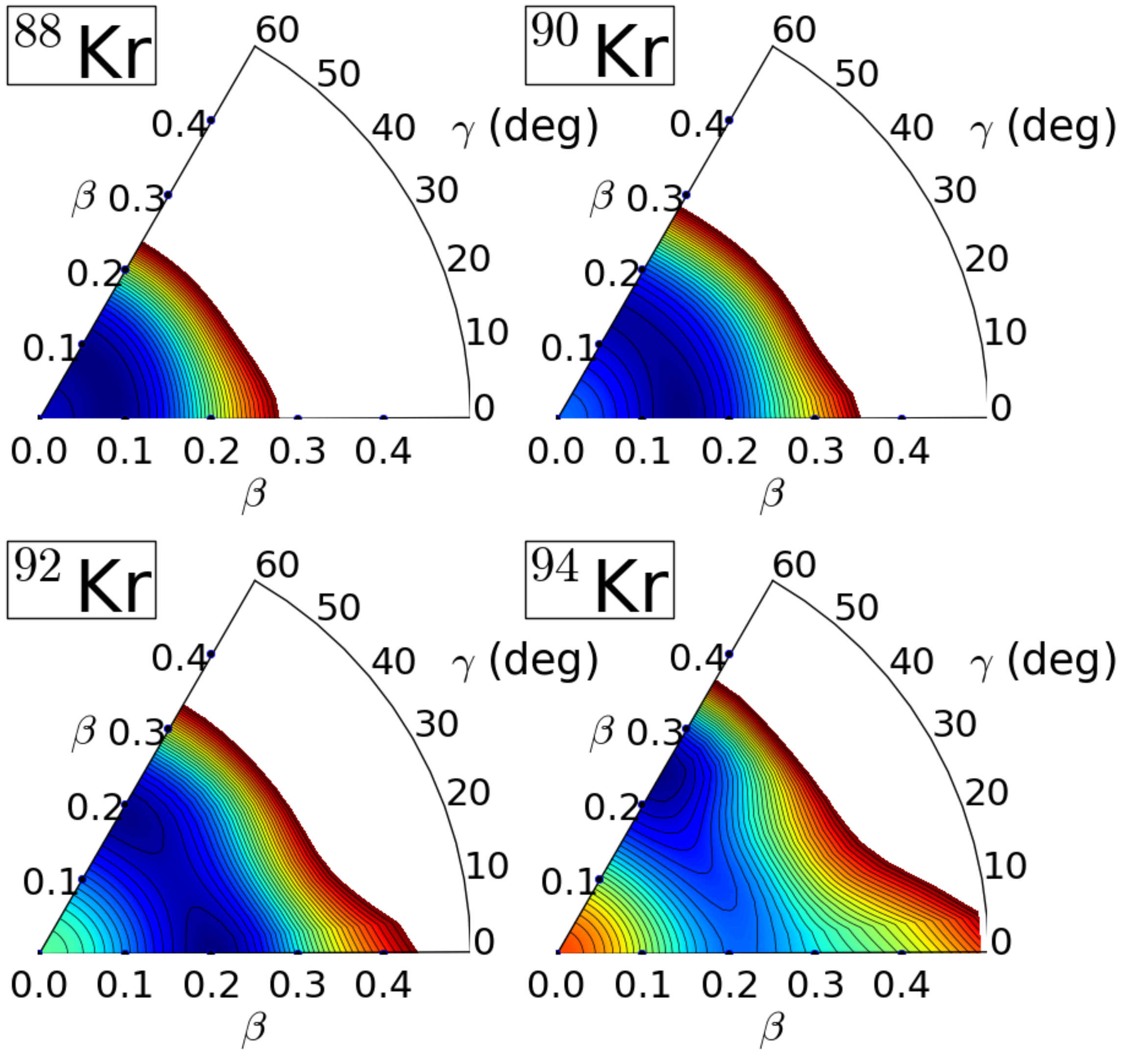}
\caption{(Color online) The Gogny-D1M $(\beta,\gamma)$-deformation
 energy surfaces obtained for the nuclei  $^{88,90,92,94}$Kr. The energy difference
 between neighboring contours is 100 keV. 
 } 
\label{fig:pes}
\end{center}
\end{figure}


\begin{figure}[htb!]
\begin{center}
\includegraphics[width=\linewidth]{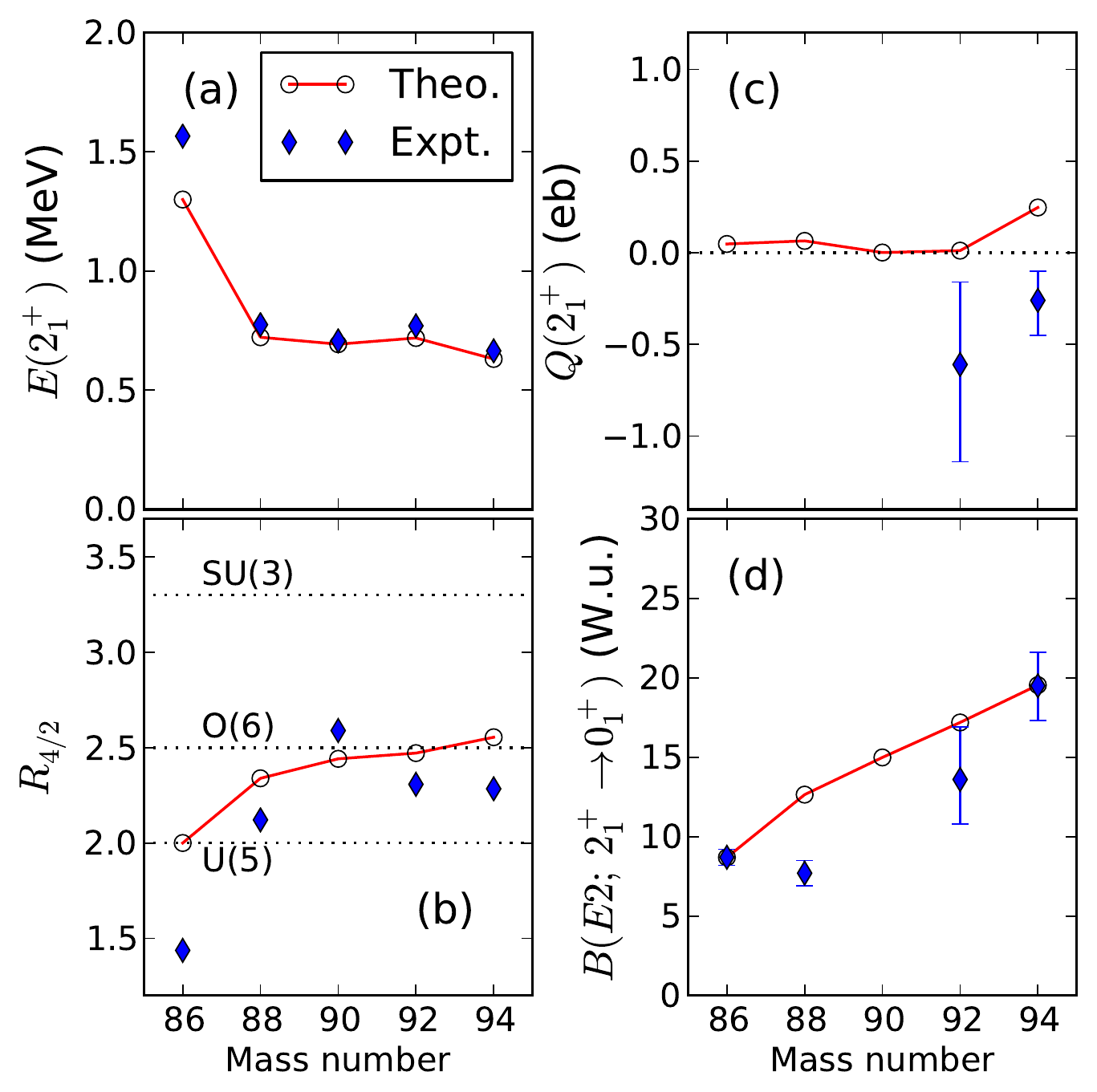}
\caption{(Color online) Evolution of the  $2^+_1$ excitation energy
 $E(2^+_1)$ [panel (a)], the $R_{4/2}$ ratio [panel (b)], the
 spectroscopic quadrupole moment $Q(2^+_1)$ [panel (c)] and 
 the $B(E2; 2^+_1\rightarrow 0^+_1)$ transition strength [panel (d)]
 as functions of the mass number for the even-even nuclei 
 $^{86-94}$Kr. For details, see the main text. 
 } 
\label{fig:even}
\end{center}
\end{figure}

In this section we briefly review some key results already obtained for 
the  even-even isotopes $^{86-94}$Kr. For a more detailed account, the 
reader is referred to Refs.~\cite{albers2012,albers2013,nomura2017kr}. 

In Figure~\ref{fig:pes} we have plotted  the $(\beta,\gamma)$  
Gogny-D1M energy surfaces obtained for $^{88-94}$Kr. In the case of 
$^{86}$Kr the energy surface, not included in the figure, exhibits a 
spherical global minimum. As can be seen from the figure, the global 
minimum evolves from nearly spherical ($^{88}$Kr) to pronounced 
$\gamma$-soft ($^{90,92}$Kr) and then to oblate ($^{94}$Kr).  As 
already mentioned, for all the nuclei depicted in the figure, no 
distinct secondary minimum has been  obtained.

In Fig.~\ref{fig:even} we have plotted the $2^+_1$ excitation energy 
$E(2^+_1)$ [panel (a)], the energy ratio $R_{4/2}=E(4^+_1)/E(2^+_1)$ 
[panel (b)], the spectroscopic quadrupole moment $Q(2^+_1)$ of the 
$2^+_1$ state [panel (c)] and the $B(E2; 2^+_1\rightarrow  0^+_1)$ 
transition strength [panel (d)] for the $^{86-94}$Kr. The experimental 
data, taken from Refs.~\cite{albers2012,albers2013,data}, are also 
included in the plots for comparisons. The  U(5) vibrational 
($R_{4/2}=2.0$), SU(3) rotational ($R_{4/2}=3.33$) and O(6) 
$\gamma$-soft ($R_{4/2}=2.5$) IBM \cite{IBM} limits are indicated in 
panel (b).
 
As can be seen from panel (a), the $2^+_1$ excitation energies 
predicted within the (mapped) IBM-2 framework nicely follow the 
experimental ones. Both the theoretical and experimental $R_{4/2}$, 
shown in panel (b), are located in between the U(5) and O(6) limits for 
most of the nuclei, and gradually increase as functions of the neutron 
number. This confirms the smooth onset of collectivity observed 
experimentally \cite{albers2012}. For the $N=50$ nucleus $^{86}$Kr, the 
experimental $R_{4/2}$ ratio is much smaller than the vibrational limit 
of 2.0. Such a feature, typical of spherical nuclei where 
single-particle dynamics is dominant, cannot be reproduced by the 
IBM-2, since it is built only on  collective pairs.

From panel (c), one sees that the IBM-2  $Q(2^+_1)$ moments are nearly 
equal to zero for all the nuclei, exception made of $^{94}$Kr. Note, 
that the Gogny-D1M energy surfaces display nearly spherical and 
$\gamma$-soft minima for $^{88}$Kr and $^{90,92}$Kr, respectively, and 
an oblate minimum for $^{94}$Kr (see, Fig.~\ref{fig:pes}). The 
predicted $Q(2^+_1)$ moments for $^{92,94}$Kr are rather at variance 
with the experimental ones \cite{albers2013} which exhibit large 
uncertainties. As can be seen from panel (d), the predicted $B(E2; 
2^+_1\rightarrow 0^+_1)$ values agree well with the experimental ones.


\section{Results for odd-mass Kr nuclei\label{sec:odd}}


In this section, we discuss the results of our calculations for the 
studied odd-mass Kr isotopes. In Sec.~\ref{systematics-ood-Kr}, we 
consider the systematic of the low-energy yrast levels. Detailed level 
schemes are discussed in Sec.~\ref{detailed-ood-Kr}. Finally, in 
Sec.~\ref{Electromag-properties}, we discuss electromagnetic properties 
such as the $B(E2)$ and $B(M1)$ transition strengths as well as  
spectroscopic quadrupole and magnetic moments. 

\subsection{Systematic of excitation energies}
\label{systematics-ood-Kr}


\begin{figure}[htb!]
\begin{center}
\includegraphics[width=0.9\linewidth]{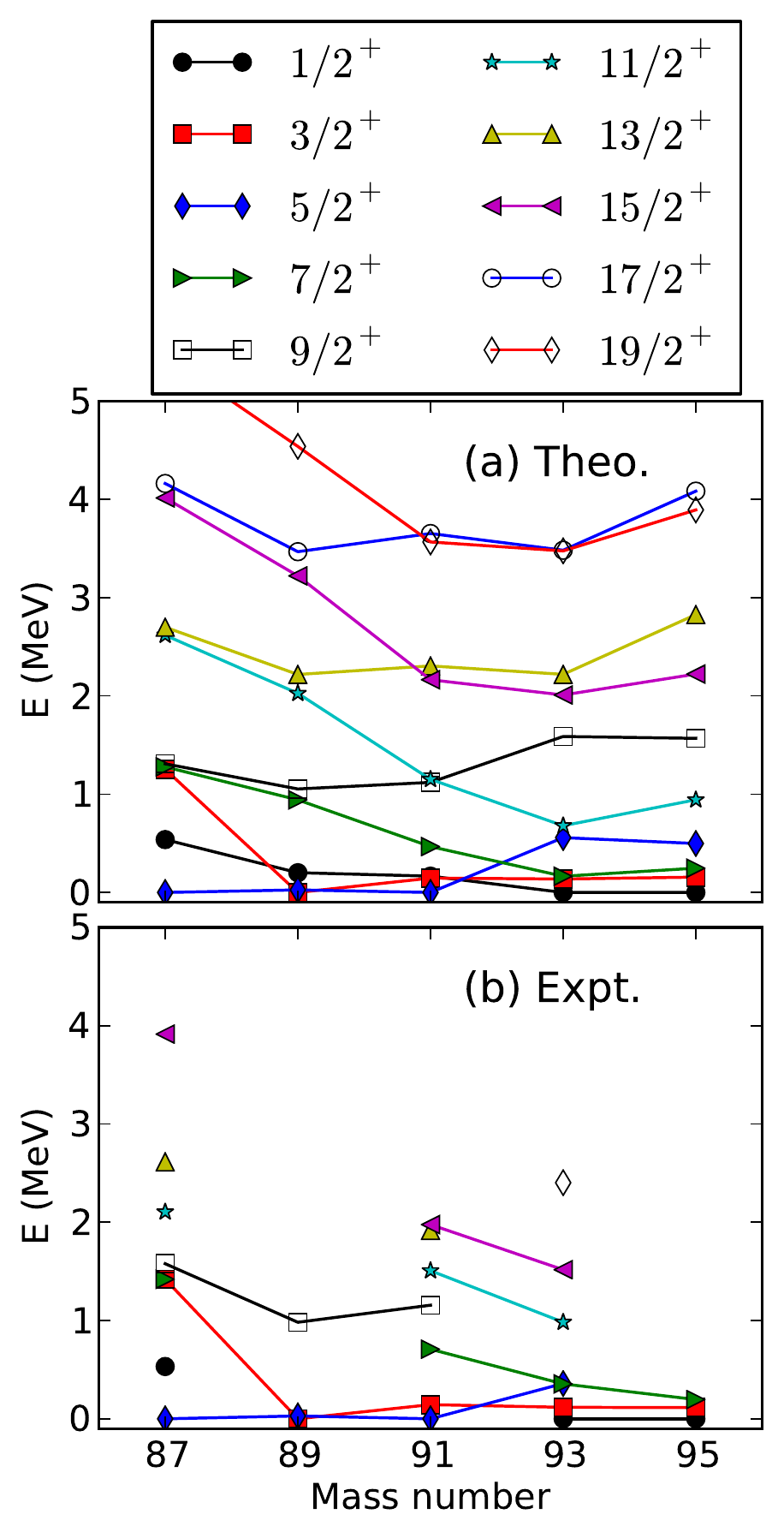}
\caption{(Color online) Evolution of the theoretical and experimental
 \cite{data,rzacaurban2017} excitation energies of the low-lying positive-parity
 yrast states in the  odd-mass nuclei  $^{87-95}$Kr as functions of the
 mass number. } 
\label{fig:odd}
\end{center}
\end{figure}

In Fig.~\ref{fig:odd} we have depicted the excitation spectra for the 
low-energy positive-parity yrast states in  $^{87-95}$Kr as functions 
of the mass number. In most cases, experimental data are only available 
for  those states in the vicinity of the ground state. As can be seen, 
the predicted spectra agree reasonably well with the experimental ones. 
From Fig.~\ref{fig:odd}, one observes a change in the predicted 
low-lying level structure between $^{91}$Kr and $^{93}$Kr. For example, 
the ${5/2}^+_1$ state is either the ground state ($^{87,91}$Kr) or in 
the vicinity of it ($^{89}$Kr). However, it goes up in energy for 
$A>91$ and  for  $^{93,95}$Kr, the state ${1/2}^+_1$  becomes the 
ground state. Note, that the neighboring even-even  isotopes around 
$A=92$ have been predicted to be in the transitional region between 
$\gamma$-soft and oblate shapes (see, Fig.~\ref{fig:pes}). As shown 
later, the IBFM-2 wave functions for the ground states of  
$^{87,89,91}$Kr  ($J={5/2}^+$ for $^{87,91}$Kr and ${3/2}^+$ for 
$^{89}$Kr) are predominantly composed of the $2d_{5/2}$ configuration 
($\approx 90$ \%) while those of $^{93,95}$Kr  with $J={1/2}^+$ are 
described by a mixture of the $2d_{3/2}$ ($\approx 50$ \%) and 
$2d_{5/2}$ ($\approx 30$ \%) configurations.

\subsection{Detailed level schemes}
\label{detailed-ood-Kr}

The detailed low-energy level schemes of the positive-parity states are 
plotted in Figs.~\ref{fig:87kr}-\ref{fig:95kr} for $^{87-95}$Kr. Both 
the experimental and theoretical spectra are shown up to around an 
excitation energy of 3 MeV. For a given spin $J$ the lowest two levels 
are plotted. The fractions of the $3s_{1/2}$, $2d_{3/2}$, $2d_{5/2}$ 
and $1g_{7/2}$ configurations in the IBFM-2 wave functions, 
corresponding to the low-lying states of the studied odd-mass nuclei, 
are given in Table~\ref{tab:wf}.


\begin{table}
 \begin{center}
\caption{\label{tab:wf} Composition of the IBFM wave functions corresponding to 
  the low-energy positive-parity states of the studied odd-mass Kr nuclei (in
  percent units).}
  \begin{tabular}{lcccccccc}
\hline\hline
 & \multicolumn{4}{c}{$^{87}$Kr} & \multicolumn{4}{c}{$^{89}$Kr} \\
\cline{2-5}
\cline{6-9}
$J$   & $3s_{1/2}$ & $2d_{3/2}$ & $2d_{5/2}$ & $1g_{7/2}$ & $3s_{1/2}$ & $2d_{3/2}$ & $2d_{5/2}$ & $1g_{7/2}$ \\
\hline
${ 1/2 }^+_{ 1}$ & 41 & 5 & 54 & 0 & 43 & 8 & 48 & 1 \\
${ 3/2 }^+_{ 1}$ & 11 & 23 & 62 & 4 & 4 & 6 & 89 & 1 \\
${ 3/2 }^+_{ 2}$ & 29 & 11 & 58 & 3 & 20 & 34 & 36 & 11 \\
${ 5/2 }^+_{ 1}$ & 4 & 1 & 95 & 0 & 6 & 3 & 90 & 1 \\
${ 5/2 }^+_{ 2}$ & 7 & 0 & 92 & 0 & 20 & 3 & 76 & 1 \\
${ 7/2 }^+_{ 1}$ & 1 & 1 & 95 & 3 & 1 & 2 & 92 & 5 \\
${ 9/2 }^+_{ 1}$ & 5 & 1 & 94 & 0 & 7 & 2 & 91 & 0 \\
${ 11/2 }^+_{ 1 }$ & 1 & 1 & 95 & 3 & 5 & 1 & 94 & 0 \\
\hline
 & \multicolumn{4}{c}{$^{91}$Kr} & \multicolumn{4}{c}{$^{93}$Kr} \\
\cline{2-5}
\cline{6-9}
$J$   & $3s_{1/2}$ & $2d_{3/2}$ & $2d_{5/2}$ & $1g_{7/2}$ & $3s_{1/2}$ & $2d_{3/2}$ & $2d_{5/2}$ & $1g_{7/2}$ \\
\hline
${ 1/2 }^+_{ 1}$ & 49 & 35 & 6 & 10 & 17 & 46 & 30 & 7 \\
${ 3/2 }^+_{ 1}$ & 7 & 56 & 6 & 31 & 58 & 21 & 10 & 11 \\
${ 3/2 }^+_{ 2}$ & 9 & 6 & 83 & 2 & 50 & 27 & 11 & 12 \\
${ 5/2 }^+_{ 1}$ & 5 & 2 & 92 & 1 & 21 & 43 & 30 & 6 \\
${ 5/2 }^+_{ 2}$ & 40 & 38 & 8 & 14 & 45 & 24 & 25 & 7 \\
${ 7/2 }^+_{ 1}$ & 4 & 37 & 0 & 59 & 60 & 19 & 5 & 16 \\
${ 9/2 }^+_{ 1}$ & 12 & 6 & 76 & 6 & 49 & 21 & 26 & 3 \\
${ 11/2 }^+_{ 1 }$ & 3 & 32 & 0 & 65 & 60 & 22 & 3 & 16 \\
\hline
 & \multicolumn{4}{c}{$^{95}$Kr} \\
\cline{2-5}
$J$   & $3s_{1/2}$ & $2d_{3/2}$ & $2d_{5/2}$ & $1g_{7/2}$ \\
\hline
${ 1/2 }^+_{ 1}$ & 11 & 48 & 36 & 5 \\
${ 3/2 }^+_{ 1}$ & 46 & 25 & 6 & 23 \\
${ 3/2 }^+_{ 2}$ & 39 & 32 & 9 & 20 \\
${ 5/2 }^+_{ 1}$ & 12 & 44 & 36 & 7 \\
${ 5/2 }^+_{ 2}$ & 42 & 21 & 20 & 17 \\
${ 7/2 }^+_{ 1}$ & 44 & 23 & 4 & 29 \\
${ 9/2 }^+_{ 1}$ & 16 & 41 & 35 & 8 \\
${ 11/2 }^+_{ 1 }$ & 42 & 24 & 2 & 32 \\
\hline\hline
  \end{tabular}
 \end{center}
\end{table}

\subsubsection{$^{87}$Kr}


\begin{figure}[htb!]
\begin{center}
\includegraphics[width=0.8\linewidth]{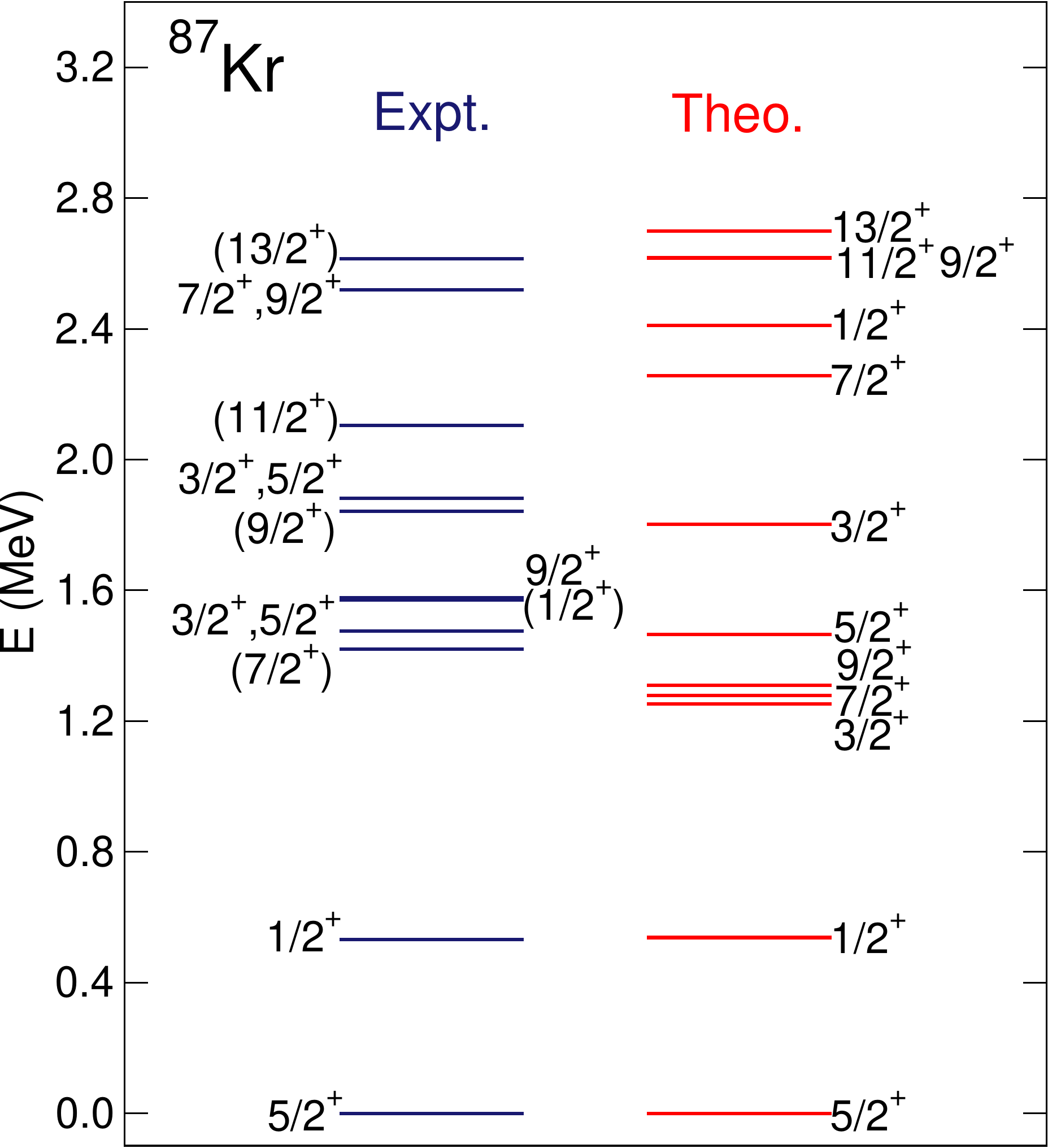}
\caption{(Color online) The low-lying
 positive-parity excitation spectrum obtained for the nucleus 
 $^{87}$Kr is compared with the available 
 experimental data Ref.~\cite{data}. Energy levels 
 in parenthesis have not been established experimentally.
 } 
\label{fig:87kr}
\end{center}
\end{figure}

Figure~\ref{fig:87kr} depicts the energy spectra for  $^{87}$Kr. The 
experimental results are well reproduced within our approach. As can be 
seen from Table~\ref{tab:wf},  95 \% of the wave function of the 
${5/2}^+_1$ ground state is accounted for by the $2d_{5/2}$ 
single-particle orbital coupled to the even-even nucleus $^{86}$Kr. For 
both, the ${1/2}^+_1$ and ${3/2}^+_1$ wave functions the $3s_{1/2}$, 
$2d_{3/2}$ and $2d_{5/2}$ configurations have a stronger mixing.

\subsubsection{$^{89}$Kr}


\begin{figure}[htb!]
\begin{center}
\includegraphics[width=0.8\linewidth]{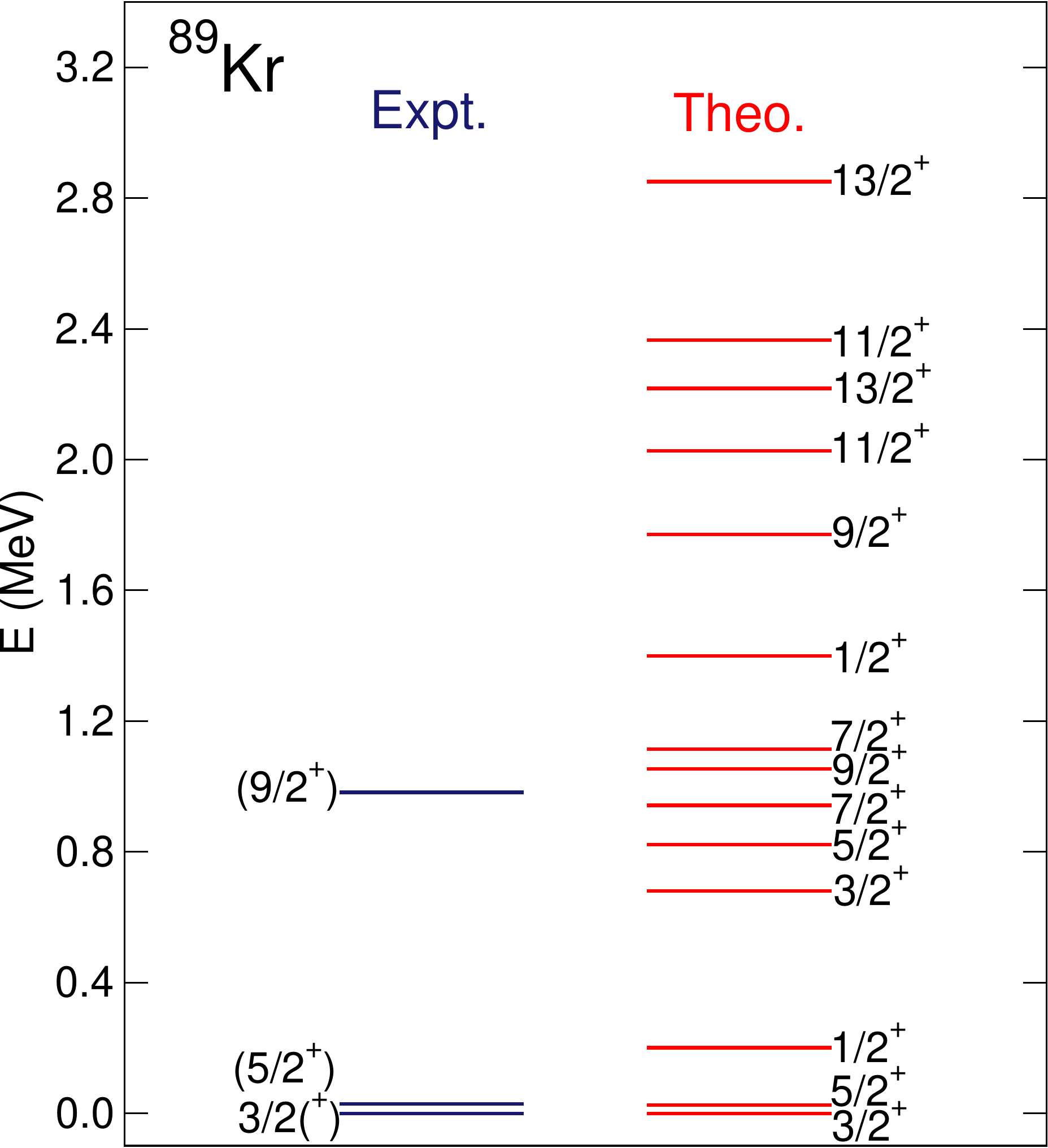}
\caption{(Color online) The same as in  Fig.~\ref{fig:87kr}, but for 
 $^{89}$Kr. Experimental data have been taken from Ref.~\cite{data}.
 } 
\label{fig:89kr}
\end{center}
\end{figure}

The experimental information is very scarce for  $^{89}$Kr. Its ground 
state has been tentatively assigned to $J^{\pi}={3/2}(^+)$ and a 
${5/2}^+$ level has been found just 29 keV above it. Moreover, this 
${5/2}^+$ state and the ${9/2}^+$ one at 982 keV, have been interpreted 
\cite{data} as the members of the $\Delta J=2$ band based on the 
neutron $2d_{5/2}$ orbital coupled to the $^{90}$Kr core.  Our 
calculations predict the ${3/2}^+_1$, ${5/2}^+_1$ and ${9/2}^+_1$ 
states to be mainly ($\approx 90\%$) composed of the $2d_{5/2}$ 
configuration (see, Table~\ref{tab:wf}). Due to the large overlaps 
between their wave functions, large E2 matrix elements have been 
obtained for the ${5/2}^+_1\rightarrow {3/2}^+_1$ and 
${9/2}^+_1\rightarrow {5/2}^+_1$ transitions [see, Figs.~\ref{fig:e2} 
panel (d) and \ref{fig:e2} panel (e)].

\subsubsection{$^{91}$Kr\label{sec:91kr}}


\begin{figure}[htb!]
\begin{center}
\includegraphics[width=0.8\linewidth]{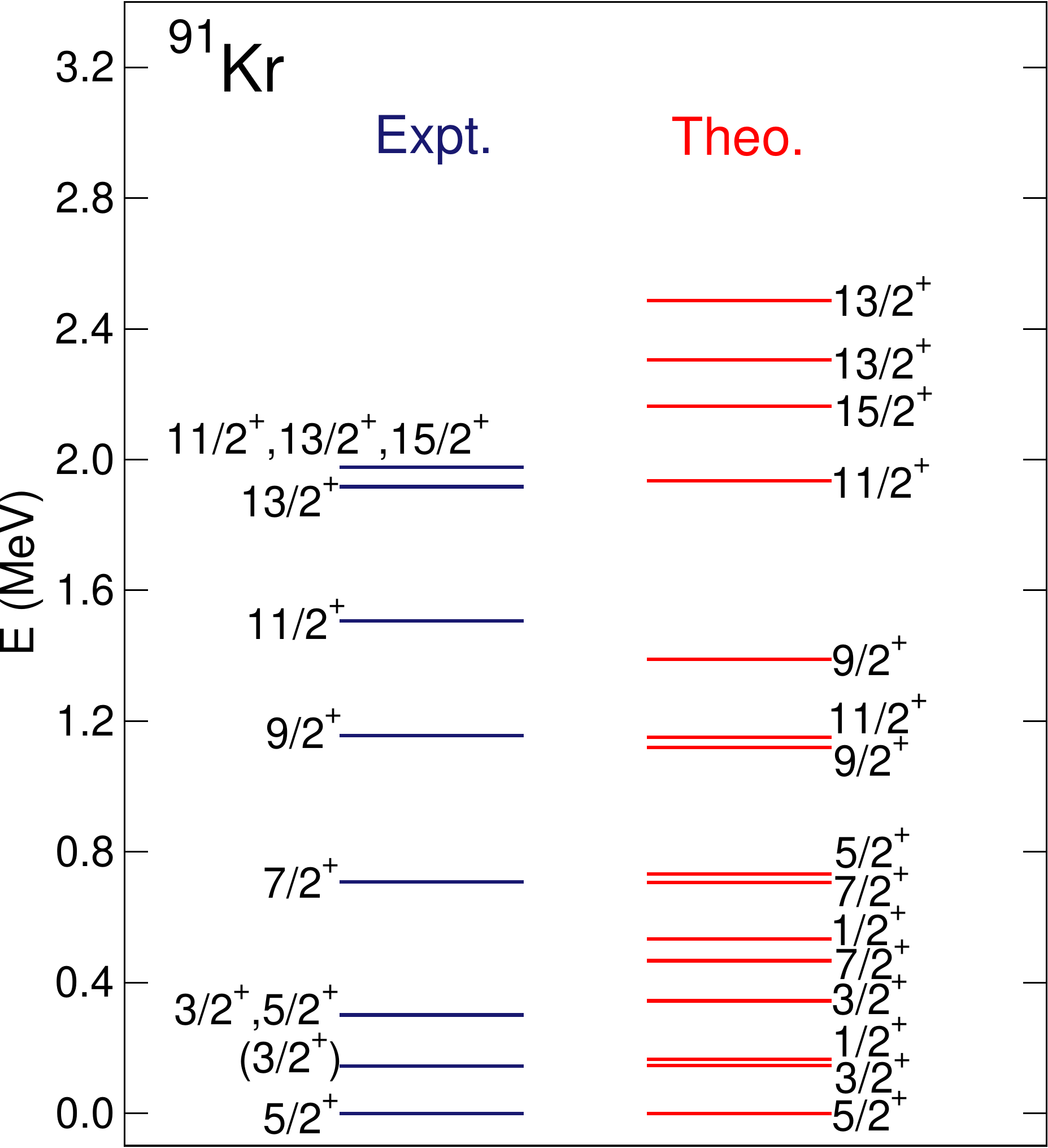}
\caption{(Color online) The same as in  Fig.~\ref{fig:87kr}, but for
 $^{91}$Kr. Experimental data have been taken from Refs.~\cite{rzacaurban2017,data}.
 } 
\label{fig:91kr}
\end{center}
\end{figure}

In the case of  $^{91}$Kr, as seen from Fig.~\ref{fig:91kr}, our 
calculations reproduce well the experimental spectrum 
\cite{rzacaurban2017,data} exception made of the ${7/2}^+_1$ and 
${11/2}^+_1$ levels which are predicted to have a rather low excitation 
energy. From Table~\ref{tab:wf}, ones observes that the wave functions 
of the ${5/2}^+_1$ ground and the ${9/2}^+_1$ states are similar,  both 
dominated by the $2d_{5/2}$ configuration. They are thus expected to be 
members of a $\Delta J=2$ band. Furthermore, the ${3/2}^+_1$ state 
(see, Table~\ref{tab:wf}) seems to be the band head of the $\Delta J=2$ 
band with the members  ${3/2}^+_1,{7/2}^+_1,\ldots$. Consequently, as 
shown later in Fig.~\ref{fig:e2}, rather large $B(E2; 
{9/2}^+_1\rightarrow {5/2}^+_1)$ and $B(E2; {7/2}^+_1\rightarrow 
{3/2}^+_1)$ transitions have been  found for $^{91}$Kr. 

In the case of $^{91}$Kr,  our calculations reproduce the experimental 
results  as well as the results of large-scale shell-model calculations 
\cite{rzacaurban2017}. However, within the shell model, neutrons 
predominantly occupy the $2d_{5/2}$ orbital for all the observed states 
in $^{91}$Kr. This is somewhat at variance with our results where, the 
${7/2}^+_1$ state is  composed of the $2d_{3/2}$ (37 \%) and $1g_{7/2}$  
(59 \%) configurations (see, Table~\ref{tab:wf}).

\subsubsection{$^{93}$Kr}


\begin{figure}[htb!]
\begin{center}
\includegraphics[width=0.8\linewidth]{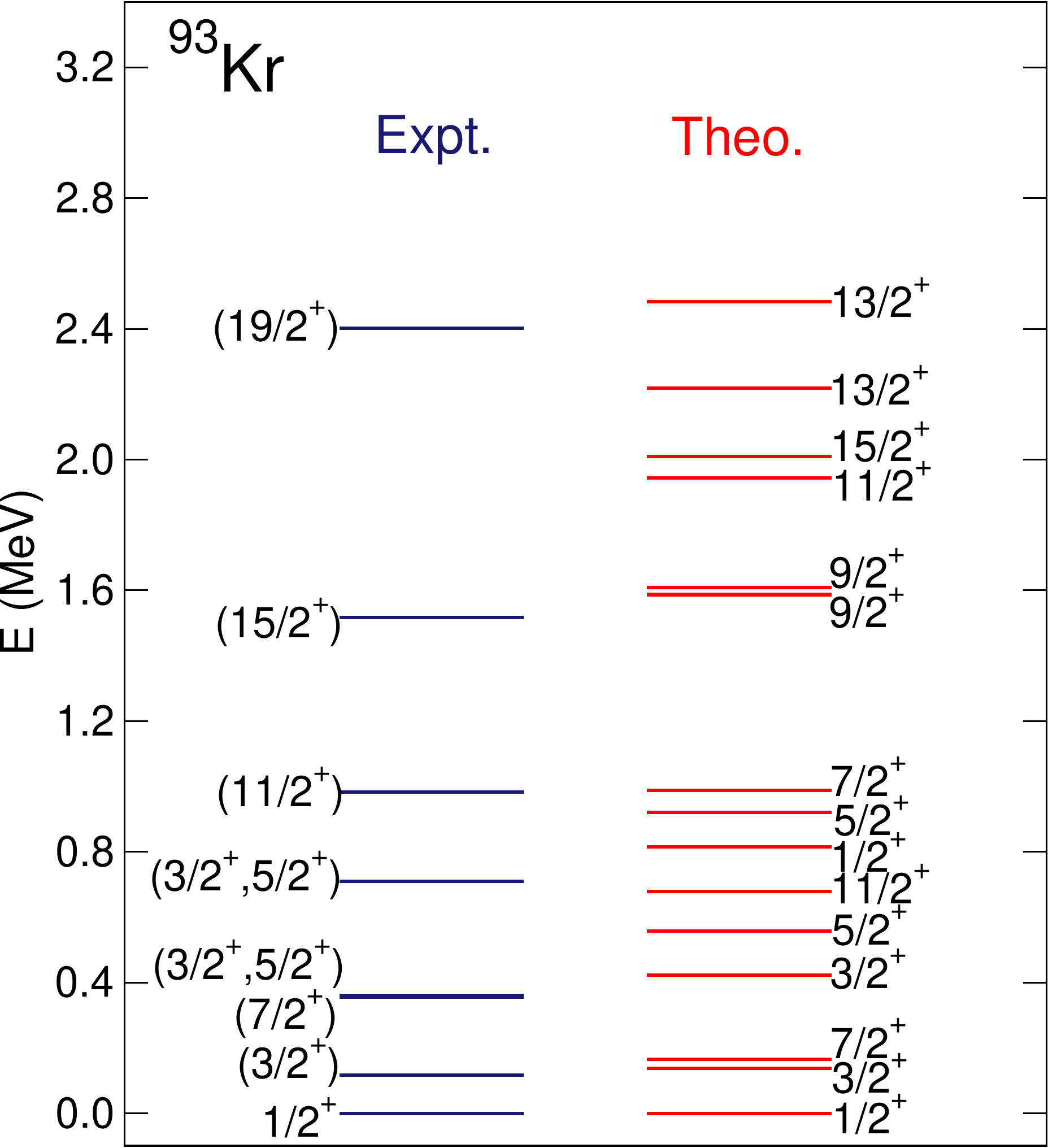}
\caption{(Color online) The same as in  Fig.~\ref{fig:87kr}, but for 
 $^{93}$Kr. Experimental data have been taken from Ref.~\cite{hwang2010,data}.
 } 
\label{fig:93kr}
\end{center}
\end{figure}

For  $^{93}$Kr, in Fig.~\ref{fig:93kr}, we conclude that the agreement 
with the experimental spectrum is reasonable. As in the case of  
$^{91}$Kr, the ${7/2}^+_1$ and ${11/2}^+_1$ levels are  too low and the 
${15/2}^+_1$ excitation energy is overestimated. In this nucleus the 
ground state corresponds to  $J={1/2}^+$. 
%
%
From the results shown in Table~\ref{tab:wf}, one realizes that in our 
calculations, the four single-particle configurations are significantly 
mixed in the $J={1/2}^+_1$ ground states of $^{93,95}$Kr. This is in 
contrast with the  $^{87-91}$Kr cases, where the ground states 
($J={5/2}^+_1$ for $^{87,91}$Kr and ${3/2}^+_1$ for $^{89}$Kr) are 
predominantly ($\approx 90\,\%$) composed of the $2d_{5/2}$ 
configuration, 

\subsubsection{$^{95}$Kr}


\begin{figure}[htb!]
\begin{center}
\includegraphics[width=0.8\linewidth]{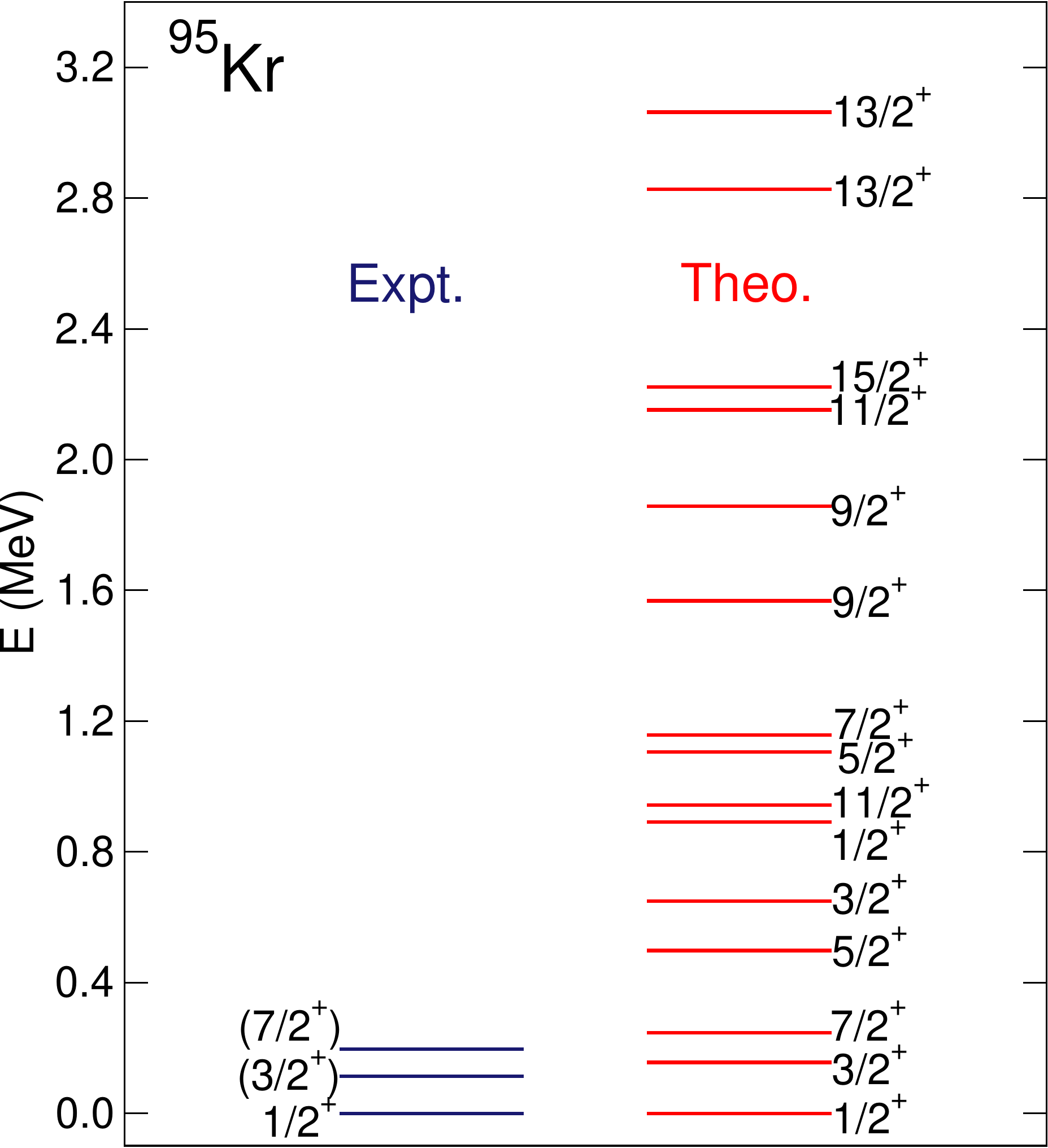}
\caption{(Color online) The same as in Fig.~\ref{fig:87kr}, but for 
 $^{95}$Kr. Experimental data have been taken from Ref.~\cite{data}.
 } 
\label{fig:95kr}
\end{center}
\end{figure}

In the case of $^{95}$Kr, the experimental information is even more 
scarce than for $^{91,93}$Kr. Here, the predicted spectrum resembles 
the one of $^{93}$Kr. From Table~\ref{tab:wf} we observe that the 
${1/2}^+_1$ ground state has a large overlap with the ${5/2}^+_1$ 
state. Both  are expected to form a $\Delta J=2$ band. On the other 
hand, the structural similarity between the ${3/2}^+_1$ and ${7/2}^+_1$ 
states suggests that they are members of another $\Delta J=2$ band.

\subsection{Electromagnetic properties}
\label{Electromag-properties}


\begin{figure}[htb!]
\begin{center}
\includegraphics[width=\linewidth]{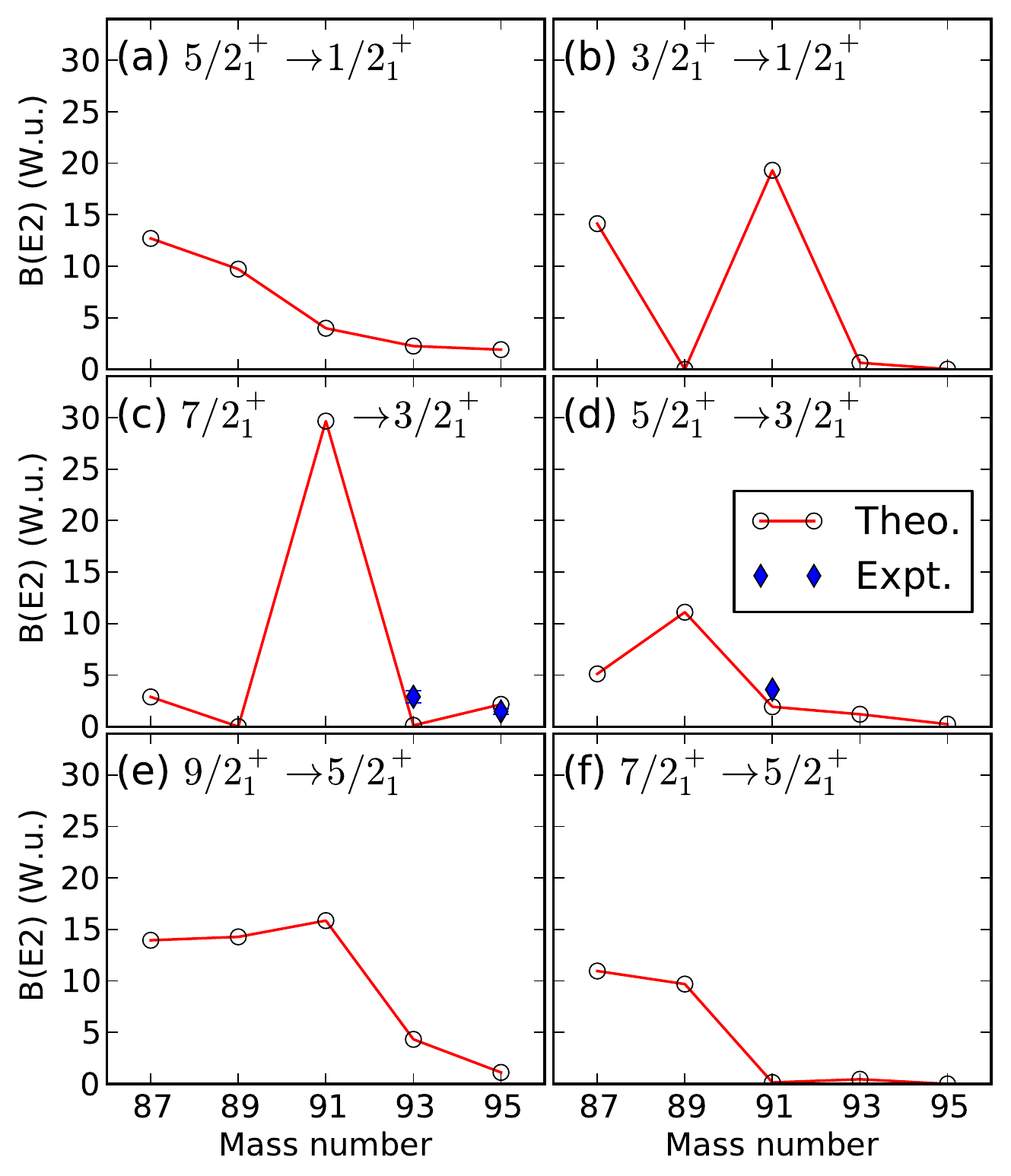}
\caption{(Color online) The $B(E2)$ transition probabilities between the low-lying
 positive-parity yrast
 states for the  odd-mass nuclei $^{87-95}$Kr are plotted, as functions of the mass
 number. The experimental data have  been taken from Ref.~\cite{data}. 
 } 
\label{fig:e2}
\end{center}
\end{figure}


\begin{figure}[htb!]
\begin{center}
\includegraphics[width=\linewidth]{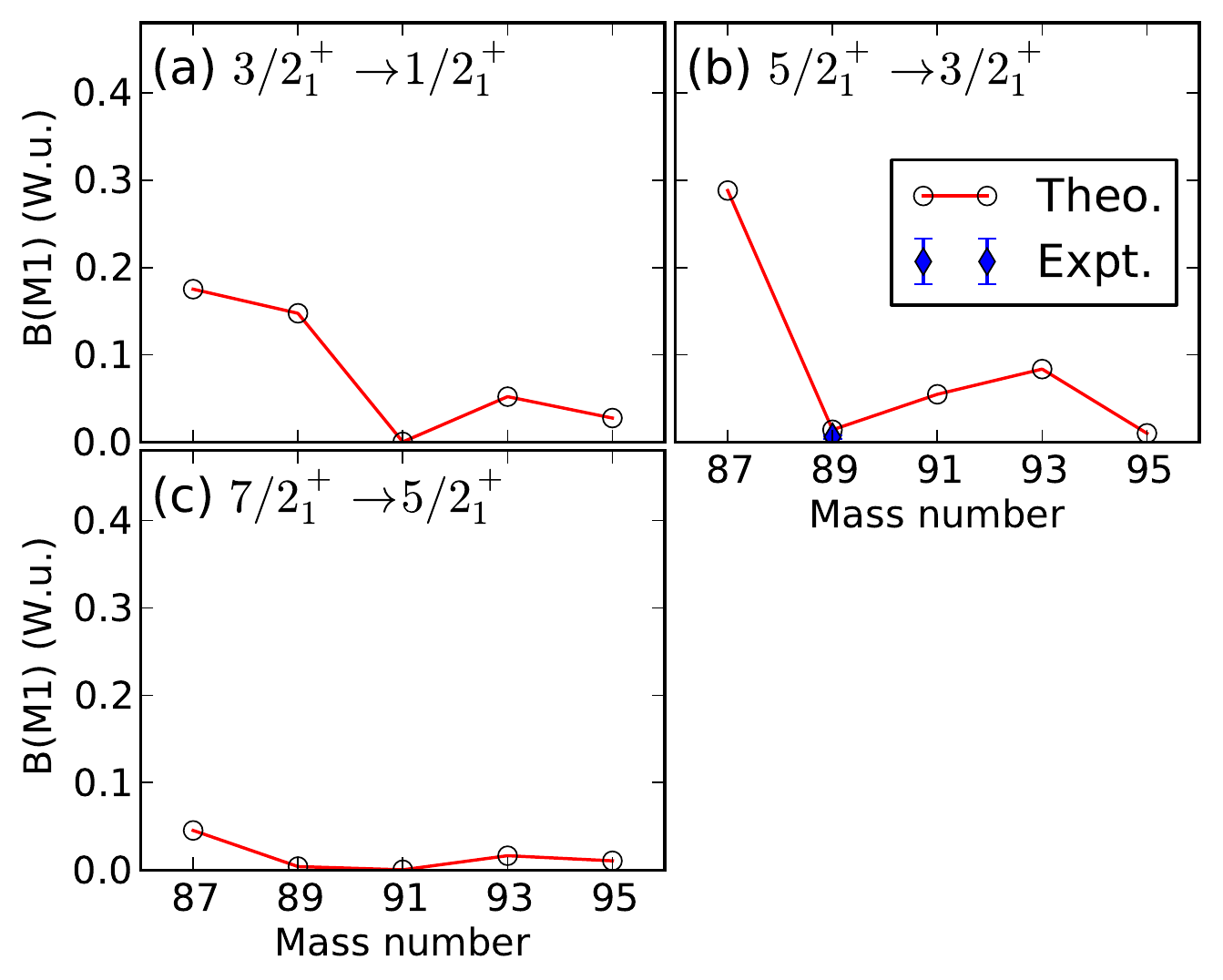}
\caption{(Color online) The same as in  Fig.~\ref{fig:e2}, but for the $B(M1)$
 rates. 
 } 
\label{fig:m1}
\end{center}
\end{figure}


\begin{figure}[htb!]
\begin{center}
\includegraphics[width=\linewidth]{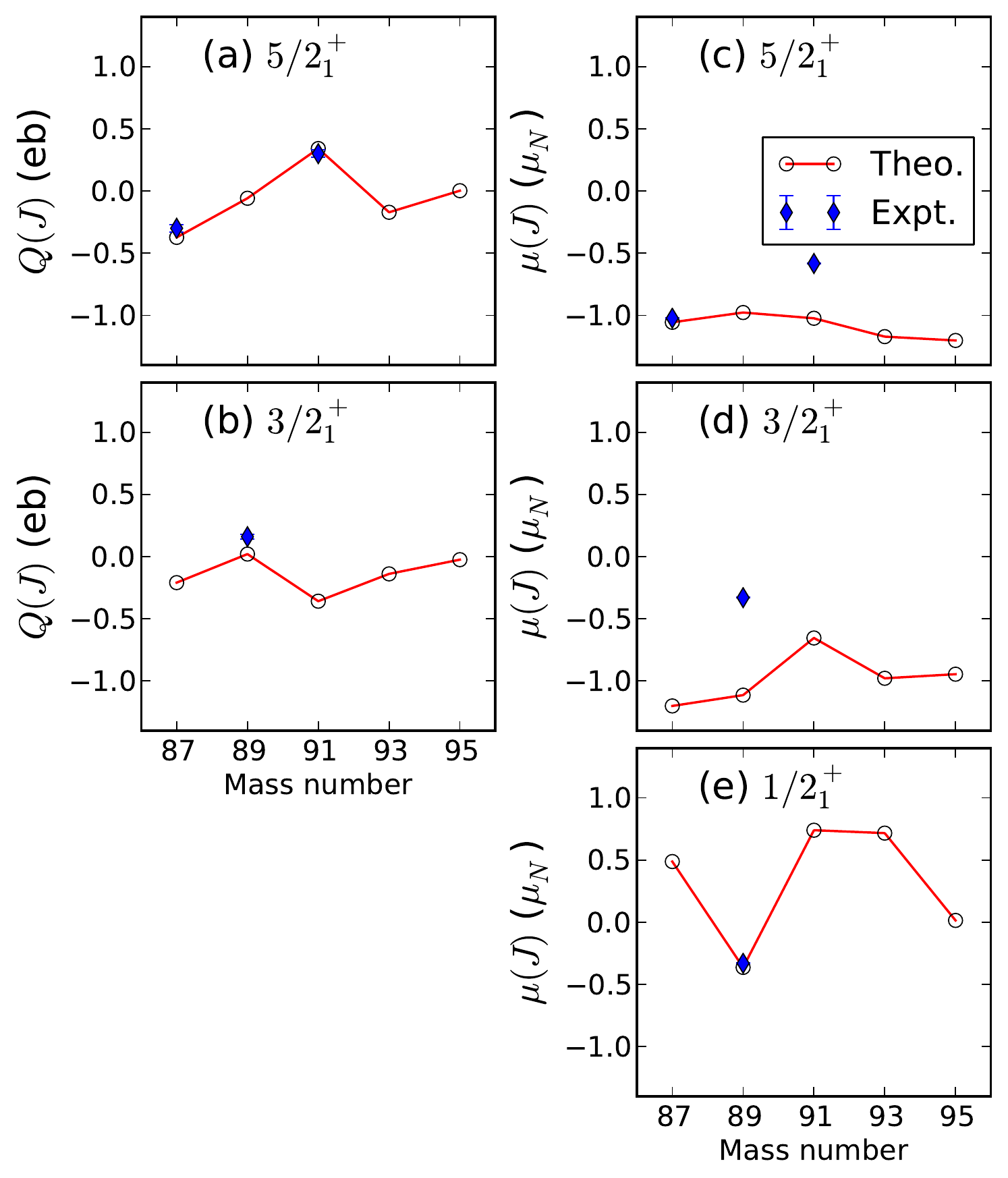}
\caption{(Color online) Spectroscopic quadrupole moments $Q(J)$ and
 magnetic moments $\mu(J)$ for the ${5/2}^+_1$, ${3/2}^+_1$ and
 ${1/2}^+_1$ states of the odd-mass isotopes $^{87-95}$Kr. The experimental
 data have been taken from Ref.~\cite{stone2005}. 
 } 
\label{fig:mom}
\end{center}
\end{figure}

The experimental information on the electromagnetic transition rates of 
the studied odd-mass Kr isotopes is rather scarce. Moreover, our choice 
of effective charges, especially for the E2 transition operator, might 
not necessarily be the optimal one. Therefore, in this section we will 
only discuss the systematic of the  $B(E2)$ and $B(M1)$ values between 
a few lowest states as functions of the nucleon number. 

Figure~\ref{fig:e2} displays the  $B(E2)$ transition probabilities for 
both the $\Delta J=1$ and 2 transitions to the ${1/2}^+_1$, 
${3/2}^+_1$, and ${5/2}^+_1$ states. Most of the computed $B(E2)$ 
values exhibit a gradual decrease as functions of the mass number. 
Nevertheless, the $B(E2; {3/2}^+_1\rightarrow {1/2}^+_1)$ and $B(E2; 
{7/2}^+_1\rightarrow {3/2}^+_1)$ values are especially large for the 
transitional  nucleus $^{91}$Kr (see, Fig.~\ref{fig:odd}). 
Especially, as mentioned in Sec.~\ref{sec:91kr}, the ${7/2}^+_1$ and
${3/2}^+_1$ states in $^{91}$Kr form the $\Delta J=2$ band that is connected by
the particularly strong $E2$ transitions in comparison to the adjacent nuclei. 
In the transitional nuclei around $^{91}$Kr, some $B(E2)$ decay patterns
should be sensitive to the detailed nuclear structure at low energy, and
the irregularity such as those observed in Figs.~\ref{fig:e2}(b) and
\ref{fig:e2}(c) could easily occur. 
The  $B(M1)$ transition rates are depicted in Fig.~\ref{fig:m1}. 

A few experimental values are available \cite{stone2005} for the 
spectroscopic quadrupole $Q(J)$ and magnetic $\mu(J)$ moments.  Those 
moments  are plotted in Fig.~\ref{fig:mom} for the ${5/2}^+_1$, 
${3/2}^+_1$ and ${1/2}^+_1$ states. In most of the cases, they exhibit 
an irregular behavior around the  $^{89}$Kr or $^{91}$Kr. Exception 
made of  $\mu({5/2}^+_1)$  for $^{91}$Kr and  $\mu({3/2}^+_1)$  for 
$^{89}$Kr, our calculations reproduce the experiment reasonably well. 


\section{Summary and concluding remarks\label{sec:summary}}


In this paper, we have studied the spectroscopic properties of 
neutron-rich odd-mass isotopes $^{87-95}$Kr within the IBFM-2 framework 
based on the Gogny-D1M EDF. The $(\beta,\gamma)$-deformation energy 
surfaces for the even-even boson-core nuclei $^{86-94}$Kr, spherical 
single-particle energies and occupation probabilities of the odd-mass 
systems have been obtained using the constrained HFB approximation. 
Those quantities have then been employed as a microscopic input to 
obtain most of the parameters of the corresponding IBFM-2 Hamiltonian. 
The remaining  parameters for the boson-fermion interaction have been 
fitted to the experimental spectrum for each odd-mass nucleus. 

The evolution of the  excitation spectra and electromagnetic properties 
for the studied odd-mass isotopes suggests a gradual change in the 
deformation of the corresponding intrinsic states. This agrees well 
with the shape/phase transition already observed in the neighboring 
even-even systems \cite{albers2012}. We have analyzed in detail the 
low-lying levels  and the corresponding wave functions for each of the 
considered  isotopes. Our results illustrate that the IBFM-2 framework, 
which has rarely been applied to this mass region, can  be used as a 
computationally feasible tool to access the spectroscopic properties of 
neutron-rich odd-mass nuclei.

We have restricted ourselves to  Kr isotopes with mass $A\leq 95$. This 
is partly because the experimental data, required to fit the 
boson-fermion parameters, are not yet available for larger mass 
numbers. On the other hand, a theoretical framework that couples 
odd-nucleon degrees of freedom to the IBM core, including intruder 
configurations associated with different intrinsic shapes, is still 
required to describe odd-mass systems beyond $A=96$ for which, shape 
coexistence is expected to play a key role. Work along these lines is 
in progress and will be reported elsewhere.

\acknowledgments
We thank Dr. A. Blazhev for valuable discussions. One of us (KN) 
acknowledges support by the QuantiXLie Centre of Excellence, a project
co-financed by the Croatian Government and European Union through the
European Regional Development Fund - the Competitiveness and Cohesion
Operational Programme (Grant KK.01.1.1.01.0004).
The  work of LMR was supported by Spanish Ministry of Economy and Competitiveness 
(MINECO) Grant Nos. FPA2015-65929-P and FIS2015-63770-P.

\bibliography{refs}

\end{document}